\pdfoutput=1
\documentclass[review]{elsarticle}
\usepackage{natbib}
\usepackage{lscape}
\usepackage[dvipdfmx]{}
\usepackage{amssymb}
\usepackage{wasysym}
\usepackage{multirow}

\journal{Nucl. Instrum. Meth. A}
\begin{document}

\begin{frontmatter}


  %%\author{Hidetoshi Otono\corref{cor1}\fnref{label2}}
  %% \ead{otono@icepp.s.u-tokyo.ac.jp}
%  \cortext[cor1]{corresponding author. E-mail : otono@phys.kyushu-u.ac.jp.}

  \title{soLenoid and tIme projectioN chAmber for neutron lifetime measurement -- LINA}


%\author[RCAPP]{H.~Otono \corref{cor1}}
\author[RCAPP]{H.~Otono}

\address[RCAPP]{Research Centre for Advanced Particle Physics, Kyushu University, Fukuoka, Japan}	
  
\begin{abstract}
Among measurements of the neutron lifetime, there is a 1\% difference between proton-counting methods and neutron-counting methods.
In this paper, a new electron-counting method with a magnetic field aiming for a 0.1\% accuracy is proposed, 
which would have a possibility to probe the discrepancy.
\end{abstract}


\end{frontmatter}

%\linenumbers

\section{Introduction}
\label{Introduction} 

According to the Particle Data Group, the neutron lifetime is reported as $880.3\pm1.1~\rm{s}$ in 2015
 %\cite{Agashe:2014kda}.
 \cite{PDG2015}.
However, there is a 1\% discrepancy, i.e., $8.4\pm2.2$ s, between two methods: 
counting surviving ultra-cold neutrons after storing ($879.6\pm0.8~\rm{s}$ %\cite{mampe1993measuring,serebrov2005measurement,Pichlmaier,Arzumanov,Steyerl}) 
\cite{Mampe,Serebrov,Pichlmaier,Arzumanov,Steyerl}) 
and counting trapped protons from the neutron decay ($888.0\pm2.1~\rm{s}$ \cite{Byrne,NIST2013}).
A proposed experiment in this paper employs an electron-counting method, based on experiments at ILL \cite{Grivot, Kossakowski,Bussiere} and
J-PARC \cite{BL05_beam,BL05_SFC, BL05_physics, BL05_BeamMonitor, BL05_TPC}.
Pulsed neutron beams pass through a time projection chamber (TPC) with a gas mixture of $\rm{^4He}:\rm{CO_2}:\rm{^3He}=85~kPa:15~kPa:100~mPa$,
which detects electrons from the neutron decay, and also measures the neutron flux with mixed ${\rm{^{3}He}}$;
A neutron captured in ${\rm {^{3}He}}$ emits a $573~\rm{keV}$ proton and a $191~\rm{keV}$ triton, 
namely the $\rm{^3He(n,p)^3H}$ reaction.
Analyzing events in the period when the neutron pulse is completely inside the TPC, the same fiducial volume can be used for both reactions.
The cold neutron pulses with a velocity of about $1000~\rm{m/s}$ have a length of approximately half the TPC.

The neutron lifetime, $\tau$, can be described with the number of the neutron decay, $N_{\beta}$, and the $\rm{^3He(n,p)^3H}$ reaction, $N_{\rm{^3He}}$, as follows:
\begin{eqnarray}
N_{\beta} &\propto& \frac{1}{\tau v},\\
N_{\rm{^3He}} &\propto& \sigma \rho  =  \sigma \times \frac{P}{T},\\
\tau &\propto& \frac{N_{\rm{^3He}}}{N_{\beta}} \frac{T}{P} \frac{1}{\sigma v},
\end{eqnarray}
where $v$ is the neutron velocity, and $\sigma$ is the cross section of the $\rm{^3He(n,p)^3H}$ reaction. 
The number density of ${\rm {^{3}He}}$ is represented as $\rho$, which depends on the pressure of ${\rm {^{3}He}}$, $P$, and the temperature
along with the neutron beam inside the fiducial volume, $T$.
%Consequently, we can obtain
%\begin{equation}
%\tau \propto \frac{N_{\rm{^3He}}}{N_{\beta}} \frac{T}{P} \frac{1}{\sigma v} = \frac{N_{\rm{^3He}}}{N_{\beta}} \frac{T}{P} \frac{1}{\sigma_0 v_0}. 
%\tau \propto \frac{N_{\rm{^3He}}}{N_{\beta}} \frac{T}{P} \frac{1}{\sigma v}. 
%\end{equation}
%Since the cross section of the $\rm{^3He(n,p)^3H}$ reaction for cold neutrons is inversely proportional to the neutron velocity, $\sigma v$ can be treated as a constant.
Since $\sigma$ for cold neutrons is inversely proportional to the neutron velocity, $\sigma v$ can be treated as a constant.

%A dominant systematic uncertainty of the experiment at ILL was related to a detection efficiency for $N_{\beta}$, i.e., $1.0\%$, 
%which would be reduced to less than $0.1\%$ with the TPC for the experiment at J-PARC.
%However, the second and third biggest systematic uncertainties still remain, which are $0.9\%$ on the subtraction of background events for $N_{\beta}$ and 
%$0.6\%$ on the separation between $N_{\beta}$ and $N_{\rm{^3He}}$, respectively.
%The proposed experiment in this paper mainly focuses on how to reduce these uncertainties.
%For other components, $P$ and $T$ were determined with an uncertainty of less than $0.5\%$ at ILL, which hopefully would be controlled within $0.1\%$ accuracy at J-PARC. 
%At $v_0 = 2200~\rm{m/s}$, $\sigma_0 = 5333 \pm7~\rm{barn} $ was obtained \cite{3He_1}, corresponding to an uncertainty of 0.13\% on the neutron lifetime.

\section{A proposed method for the neutron lifetime experiment}

\begin{figure}[htbp]
  \begin{center}
  \centering
  \includegraphics[width=85mm,bb=0 0 900 450]{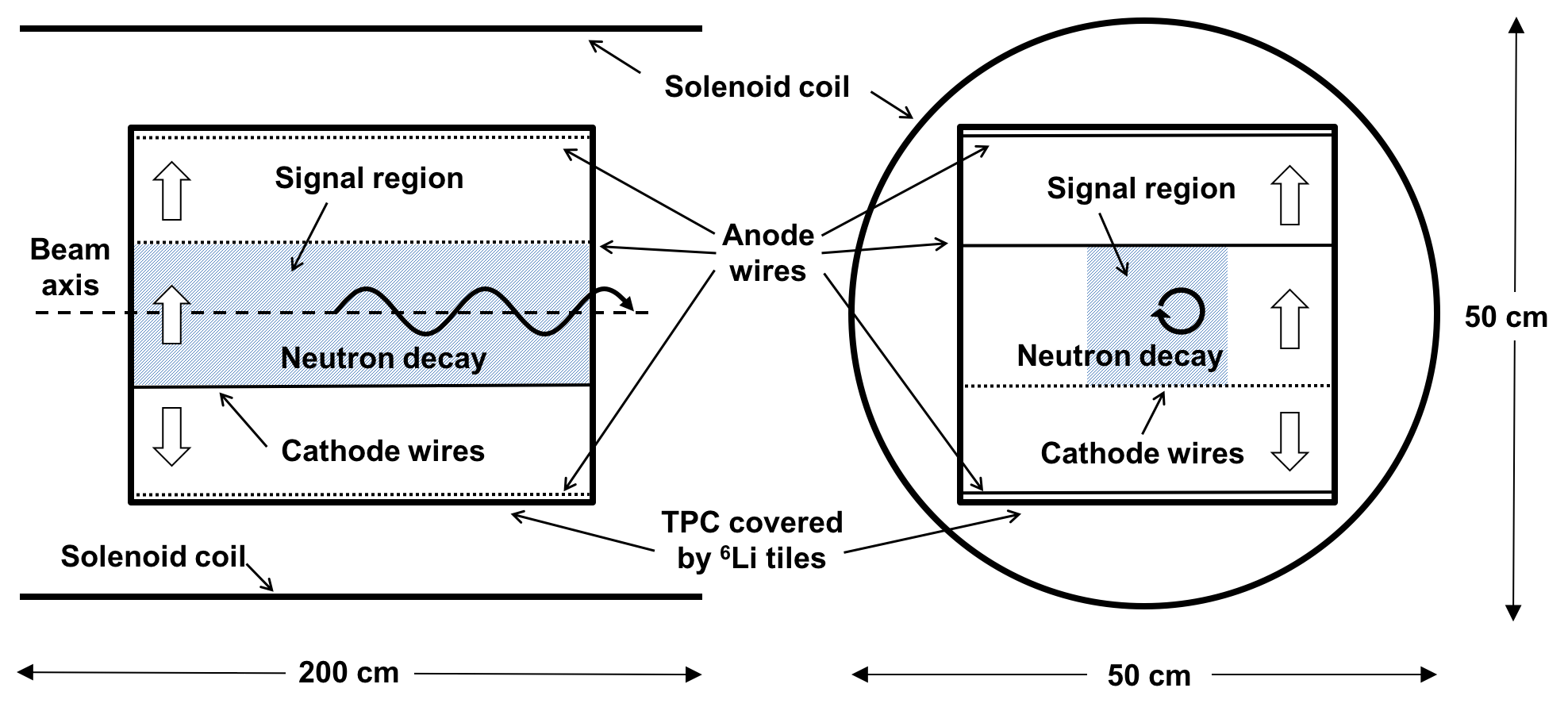}
  \end{center}
  \caption{Side view (left) and front view (right) of the proposed experiment. Not to scale. Anode (cathode) wires are assumed to be perpendicular (parallel) to the beam axis.
  	Arrows shows drift directions.}
  \label{fig:LINA}
\end{figure}

%New features of the proposed method are to employ an uniform magnetic field and an unique electric field for a TPC.  
%\subsection{Concept}
Figure \ref{fig:LINA} shows a schematic view of the proposed experiment, where a TPC is housed in a solenoid coil.
To consider the setup, I refered to the solenoid magnet for the {\sc Perkeo} $\rm{I\hspace{-1pt}I\hspace{-1pt}I}$ experiment \cite{Perkeo3};
The operation field is $1.5~{\rm{kG}}$;
The dimension for the inner surface of the magnet is $50~{\rm{cm}~(\phi)} \times 200~{\rm{cm}~(L)}$.
%The detector space within $\pm 0.5\%$ field tolerance is $50~{\rm{cm}~(L)} \times 60~{\rm{cm}~(\phi)}$.

%\vspace*{10mm}

%\vspace*{10mm}

In a magnetic field of $B~\rm{[kG]}$, 
the radius of curvature of $R~\rm{[cm]}$ for a charged particle with a mass of $m~\rm{[MeV/c^2]}$ can be denoted as follows:
\begin{equation}
R = \frac{p}{0.3B} = \frac{\sqrt{T^{2}+2mT}}{0.3B},
\end{equation}
where $p~\rm{[MeV/c]}$ and $T~\rm{[MeV]}$ are momentum and kinetic energy of the charged particle.
For a $782~\rm{keV}$ electron, corresponding to the maximum kinetic energy of electrons from the neutron decay, 
the radius of curvature is calculated to be $2.6~\rm{cm}$.
The TPC is divided into three regions by anode and cathode wires, and the drift direction in each region is indicated by arrows.
A $B \times E$ effect on the drift direction is about $3^{\circ}$, assuming an electric field of $300~\rm{V/cm}$. 
%Only trajectories of charged particles inside the central region can be transported to the anode wires,
Thus, electrons from the neutron decay and their ionised electrons are confined in the signal region shown as the shadow area.

%\subsection{Rejection of background events for the neutron decay}

Neutrons passing through the TPC are sometimes scattered by gases and captured in the structure of the detector system.
Then, prompt $\gamma$-rays are produced via the (n,$\gamma$) reaction and induce Compton electrons inside the TPC, which are irreducible background events for the neutron decay.
The surface of the TPCs for the experiments at ILL and J-PARC were covered by $\rm{^{6}Li}$ tiles, since $\rm{^{6}Li}$ has a large neutron absorption cross section without emitting the prompt $\gamma$-rays.
However, a $0.9\%$ uncertainty on the subtraction of the Compton electrons was still remained, according to the result at ILL {\it{et al.}} \cite{Kossakowski}.
For the experiment in J-PARC, this uncertainty would be dominant, 
since the other systematic uncertainties would be suppressed \cite{BL05_TPC}.
Note that these two TPCs had a single drift direction without a magnetic field.

%Since the kinetic energy of the prompt $\gamma$-rays from the $\rm{^{6}Li}$ tiles is below $7.2~\rm{MeV}$,
%the resulting Compton electrons have the radius of curvature of less than $10~\rm{cm}$ in a magnetic filed of $2.5~{\rm{kG}}$.
The proposed TPC having the three regions with the solenoid coil
would eliminate the Compton electrons by requiring no hits outside the signal region,
since the Compton electrons are mainly originated from the structure of the TPC, e.g., the $\rm{^{6}Li}$ tiles.
%The anodes and cathodes are located $10~\rm{cm}$ away from the top and bottom surfaces of the TPC;
%The events with a hit within $10~\rm{cm}$ of the longer side surfaces of the TPC are removed.
As a result, the uncertainty on the subtraction of the Compton electrons would be negligible.

\section{Conclusion}

A new electron-counting method for the neutron lifetime is proposed.
The TPC counts electrons from neutron decay and the $\rm{^3He(n,p)^3H}$ reaction for the neutron flux in the same fiducial volume. 
The current main systematic uncertainties are related to the subtraction of background events against electrons from the neutron decay.
The newly introduced magnetic and electric fields for the TPC significantly reduce the uncertainty,
which would reach a 0.1\% accuracy and offer a clue to help resolve the 1\% discrepancy among the current neutron lifetime measurements.

%\bibliography{ref.bib}

\vspace*{10mm}



 \begin{thebibliography}{00}

\bibitem{PDG2015} K.~A.~Olive {\it{et al.}}, Chin. Phys. C, {\bf{38}}, 090001 (2014) and 2015 update.
\bibitem{Mampe} W.~Mampe {\it{et al.}}, JETP Lett. {\bf{57}}, 82 (1993).
\bibitem{Serebrov} A.~P.~Serebrov {\it{et al.}}, Phys. Lett. B {\bf{605}}, 72 (2005).
\bibitem{Pichlmaier} A.~Pichlmaier {\it{et al.}}, Phys. Lett. B {\bf{693}}, 221 (2010).
\bibitem{Arzumanov} S.~S.~Arzumanov {\it{et al.}}, JETP Lett. {\bf{95}}, 224 (2012).
\bibitem{Steyerl} A.~Steyerl {\it{et al.}}, Phys. Rev. C {\bf{85}}, 065503 (2012).
\bibitem{Byrne} J.~Byrne {\it{et al.}}, Europhys. Lett. {\bf{33}}, 187 (1996).
\bibitem{NIST2013} A.~T.~Yue {\it{et al.}}, Phys. Rev. Lett. {\bf{111}}, 222501 (2013).
\bibitem{Grivot} P.~Grivot {\it{et al.}}, Nucl. Instr. and Meth. A {\bf{34}}, 127-134 (1988).
\bibitem{Kossakowski} R.~Kossakowski {\it{et al.}}, Nucl. Phys. A, {\bf{503}}, 473-500 (1989).
\bibitem{Bussiere} A.~Bussi\'ere {\it{et al.}}, Nucl. Instr. and Meth. A {\bf{332}}, 220-223 (1993).
\bibitem{BL05_beam} K.~Mishima {\it{et al.}}, Nucl. Instr. and Meth. A {\bf{600}}, 342-345 (2009).
\bibitem{BL05_SFC} K.~Taketani {\it{et al.}}, Nucl. Instr. and Meth. A {\bf{634}}, S134-S137 (2011).
\bibitem{BL05_physics} Y.~Arimoto {\it{et al.}}, Prog. Theor. Exp. Phys., 02B007 (2012).
\bibitem{BL05_BeamMonitor} T.~Ino {\it{et al.}}, J. Phys. Conf. Ser. {\bf{528}}, 012039 (2014).
\bibitem{BL05_TPC} Y.~Arimoto {\it{et al.}}, Nucl. Instr. and Meth. A {\bf{799}}, 187-196 (2015).
\bibitem{Perkeo3} B.~M\"arkisch {\it{et al.}}, Nucl. Instr. and Meth. A {\bf{611}}, 216-218 (2009).   
\end{thebibliography}
\end{document}